\documentclass[12pt]{iopart}
\usepackage{iopams}
\usepackage{graphicx}
\newcommand{\veps}{\varepsilon}
\newcommand{\rhm}{\rho_{\mathrm{m}}}
\newcommand{\prm}{p_{\mathrm{m}}}
\newcommand{\dph}{\dot{\phi}}

\begin{document}
\title{To theory of gravitational interaction}

\author{A.V. Minkevich}

\address{Department of Theoretical Physics, Belarussian State University, Belarus}
\address{Department of Physics and Computer Methods, Warmia and Mazury University in Olsztyn,
Poland}
\eads{minkav@bsu.by, awm@matman.uwm.edu.pl}

\begin{abstract}
Some principal problems of general relativity theory and attempts
of their solution are discussed. The Poincar\'e gauge theory of
gravity as natural generalization of Einsteinian gravitation
theory is considered. The changes of gravitational interaction in
the frame of  this theory leading to the solution of principal
problems of general relativity theory are analyzed.
\end{abstract}

\section{Introduction}

Einsteinian general relativity theory (GR) is the base of modern
theory of gravitational interaction, relativistic cosmology and
astrophysics. GR allows to describe different gravitating systems
and cosmological models at widely changing scales of physical
parameters. At the same time GR possesses certain principal
difficulties, which, in particular, appear in cosmology. One of
the most principal cosmological problems remains the problem of
cosmological singularity (PCS): various cosmological models
describing the evolution of the Universe have the beginning in
time in the past and in accordance with Einstein gravitation
equations the singular state with divergent energy density and
singular metrics appears at the beginning of cosmological
expansion. The PCS is particular case of general problem of
gravitational singularities of GR appearing by description of
gravitating systems at extreme conditions (extremely high energy
densities and pressures) [1]. Other principal problem of GR is
connected with explanation of cosmological and astrophysical
observations. To explain observational cosmological and
astrophysical data in the framework of GR it is necessary to
suppose that approximately 96\% energy in the Universe is related
to some hypothetical kinds of gravitating matter -- dark energy
and non-baryonic dark matter, and only 4\% energy is related to
usual gravitating matter, from which galaxies are built. As a
result, the situation in cosmology and generally in gravitation
theory actually in certain relation is similar to that in physics
at the beginning of XX century, when the notion of "ether" was
introduced in order to explain various electrodynamic phenomena.
As it is well known, the creation of special relativity theory by
A. Einstein allowed to solve existed problems without "ether"
notion.

There were many attempts with the purpose to solve indicated
problems of GR. We will discuss briefly the most known from such
attempts. Because in the frame of GR there is not restrictions on
admissible values of  energy density, and  the energy density can
reaches the Planckian scale, according to opinion of many
physicists the solution of PCS has to be connected with quantum
gravitation theory. A number of regular cosmological solutions was
obtained in the frame of candidates to quantum gravitation theory
- string theory/M-theory and loop quantum gravity [2 - 5]. Radical
ideas connected with notions of strings, branes, extra-dimensions,
space-time foam etc are used in these works. Note that indicated
works are not free from some problems and difficulties. So, the
obtaining of regular cosmological solutions in the frame of string
theory is connected with breakdown of physical condition of energy
density positivity for gravitating matter (see, for example, [2,
3]). Moreover, the most part of cosmological solutions in string
theory by transition to 4-dimensional system of reference are
singular [6]. In connection with this note that the solution of
PCS, from our point of view, means not only  obtaining regular
cosmological solutions, but also excluding singular solutions,
this means all physically acceptable cosmological solutions (or
the most part of such solutions) in the frame of correct
gravitation theory have to be regular. Bouncing cosmological
solutions obtained in loop quantum gravity [4, 5] by more exact
calculations contain gravitational singularity with divergent
Hubble parameter [7]. The dark energy or quintessence as
hypothetic kind of gravitating matter with negative pressure was
introduced with the purpose to explain accelerating cosmological
expansion at present epoch in the frame of GR [8]. In many papers
the dark energy is related to vacuum energy leading to
cosmological constant in Einstein gravitation equations. By taking
into account that the vacuum energy density has divergent value in
quantum field theory and can be eliminated by means of
regularization procedure, the following question appears: why only
the very small part of it is manifested as cosmological constant?
Unlike dark energy, the distribution of dark matter in space is
not homogeneous; by using the method of gravitational lensing,
dark matter maps of its distribution in the Universe were made
[9]. The nature of dark matter is unknown still now; according to
opinion of many physicists the dark matter is formed from weak
interacting massive particles (WIMP) revealed in united
supersymmetric models of elementary particles.

There is other treatment with the purpose of the solution of
discussed problems by using non-Einsteinian theories of gravity.
So, now one discusses largely gravitation theory in Riemannian
space-time, based on gravitational Lagrangian in the form of some
function of a scalar curvature with the purpose to solve the dark
energy problem of GR [10]. The phenomenological change of Newton's
dynamics was considered in so-called MOND in order to solve the
dark matter problem [11]. In the frame of different
non-Einsteinian theories of gravity the PCS was also studied. We
do not have an aim to consider here all these theories. Note only
that the most part of various generalizations of  Einsteinian
theory of gravitation do not have solid theoretical foundation.

At the same time now there is the gravitation theory built in the
framework of common field-theoretical approach including the local
gauge invariance principle in 4-dimensional physical space-time,
which is natural generalization of GR and which offers
opportunities to solve its principal problems. It is the
Poincar\'e gauge theory of gravity (PGTG). The main goal of this
paper is to attract attention to this fact. In Section 2 the
question "Why we need the PGTG?" is considered. In Section 3 the
changes of law of gravitational interaction by certain physical
conditions in the frame of PGTG and their physical consequences
are discussed.

\section{Why we need the Poincar\'e gauge theory of gravity?}

As it is known, the local gauge invariance principle is the basis
of modern theories of fundamental physical interactions. From
physical point of view, this principle establishes the
correspondence between certain important conserving physical
quantities, connected according to the Noether theorem with some
symmetries groups, and fundamental physical fields, which have as
a source corresponding physical quantities and play the role of
carriers of fundamental physical interactions. The applying of
this principle to gravitational interaction leads, generally
speaking, to generalization of Einsteinian theory of gravitation.
Note that because the sources of gravitational field are connected
with space-time transformations, the gauge treatment to
gravitation has some differences in comparison with Yang-Mills
fields connected with internal symmetries groups.

GR and generally metric theories of gravity, in the frame of which
the energy-momentum tensor plays the role of source of
gravitational field, can be introduced by localizing the
4-parametric translations group [12, 13]. Because the localized
translations group leads us to general coordinate transformations,
from this point of view the general covariance of GR is connected
with gauge approach. At the same time the local Lorentz group
(group of tetrad Lorentz transformations) in GR and other metric
theories of gravitation does not play any dynamical role from the
point of view of gauge approach, because the corresponding Noether
invariant in these theories is identically equal to zero [14]. The
including of the tetrad Lorentz group to gravitational gauge group
leads to the PGTG [15] (see also e.g. [16] and references herein).
In the frame of PGTG the gauge Lorentz field, which has
transformation properties of anholonomic Lorentz connection [17],
is considered together with orthonormal tetrad as independent
gravitational field variables, as a result the PGTG is gravitation
theory in the 4-dimensional Riemann-Cartan space-time $U_{4}$. By
other words, if one means that the Lorentz group, which is
fundamental group in physics, plays the dynamical role in the
gauge field theory and the Lorentz gauge field exists in the
nature, we obtain necessarily the gravitation theory in the
Riemann-Cartan space-time as natural generalization of GR. The
torsion tensor $S^i{}_{\mu\nu}$  and the curvature tensor
$F^{ik}{}_{\mu\nu}$  play the role of gravitational field
strengths in PGTG and are defined by the tetrad $h^i{}_\mu$ and
the Lorentz connection $A^{ik}{}_\mu$ by the following way:
\begin{eqnarray}\label{1}
 S^i{}_{\mu \,\nu }  = \partial _{[\nu } \,h^i{}_{\mu ]} -
h_{k[\mu } A^{ik}{}_{\nu ]}\,,  \quad F^{ik}{}_{\mu\nu } =
2\partial _{[\mu } A^{ik}{}_{\nu ]}  + 2A^{il}{}_{[\mu }
A^k{}_{|l\,|\nu ]}\,,
\end{eqnarray}
where holonomic and anholonomic components are denoted by means of
greek and latin indices respectively. If one uses minimal coupling
of matter with gravitational field, the energy-momentum and spin
tensors play the role of sources of gravitational field in PGTG.
Unlike gauge Yang-Mills fields, the translational gauge strength
-- the torsion tensor defined by (1) -- depends also on the
Lorentz gauge field. As a result in general case the torsion can
be created by the energy-momentum as well as by the spin momentum
of gravitating matter. Despite the opinion presented in literature
(see, for example, [18]) that the torsion (non-Riemannian
space-time characteristics) is essential only for gravitating
matter having the spin momentum, really the torsion can play the
principal role in the case of usual spinless gravitating systems
(see below Section 3). Although the direct interaction of the
torsion with minimally coupled spinless matter is absent, the
dynamics of such gravitating systems depends essentially on
space-time torsion by virtue of the interaction between metric and
torsion fields. Note also that the correspondence between
conserving physical quantities and gauge fields discussed at the
beginning of this Section has some peculiarity in the case of
PGTG. The law of conservation of angular momentum current in the
frame of the field theory in Minkowsky space-time includes
together with the spin also orbital part. The procedure leading to
this conservation law is realized by using the system of
orthogonal cartesian coordinates, where the coordinate basis and
the local Lorentz tetrad coincide, and the origin of orbital part
of angular momentum current is connected with coordinate
transformations essentially and its tensor definition is possible
only in this case. However, the orbital part of angular momentum
current is not source of gravitational field in PGTG; really it
can not play this role in the frame of not homogeneous space-time,
where its tensor definition does not exist (cf. [19]). Note also
that only localized tensor quantities can be sources of physical
fields. Unlike spin momentum the angular orbital momentum does not
have this property: even in Minkowsky space-time, where tensor
definition  of angular orbital momentum is possible, its value
depends on the choice of the origin of coordinates system. We will
consider below the PGTG as relativistic gravitation theory in
4-dimensional Riemann-Cartan space-time, in the frame of which the
gravitational field is described by means of interacting metric
and torsion fields and is created by energy-momentum and spin
tensors of gravitating matter.

The structure of gravitational equations of PGTG depends on the
choice of gravitational Lagrangian ${\cal L}_{\rm g}$ built by
means of gravitational field strengths (and tetrad or metrics).
The simplest PGTG is the Einstein-Cartan theory, which corresponds
to the choice of ${\cal L}_{\rm g}$ in the form of scalar
curvature of $U_{4}$ and in the frame of which the torsion and
spin tensors are connected by linear algebraic way [15, 20]. The
Einstein-Cartan theory is degenerate theory; in particular, in the
frame of this theory the torsion is equal to zero identically for
spinless matter, although the torsion tensor, as it was noted
above, is gravitational strength corresponding to translations
connected with energy-momentum tensor directly. Similar to
Yang-Mills fields the gravitational Lagrangian of PGTG has to
include terms quadratic in gravitational field strengths. Because
explicit form of quadratic part of ${\cal L}_{\rm g}$ is unknown,
we will consider PGTG by choosing ${\cal L}_{\rm g}$ in general
form including various invariants quadratic in the curvature and
torsion tensors:
\begin{eqnarray}\label{2}
{\cal L}_{\rm g}=  f_0\,
F+F^{\alpha\beta\mu\nu}\left(f_1\:F_{\alpha\beta\mu\nu}+f_2\:
F_{\alpha\mu\beta\nu}+f_3\:F_{\mu\nu\alpha\beta}\right)+
F^{\mu\nu}\left(f_4\:F_{\mu\nu}+f_5\:
F_{\nu\mu}\right)\nonumber \\  + f_6\:F^2
+S^{\alpha\mu\nu}\left(a_1\:S_{\alpha\mu\nu}+a_2\:
S_{\nu\mu\alpha}\right)
+a_3\:S^\alpha{}_{\mu\alpha}S_\beta{}^{\mu\beta}, 
\end{eqnarray}
where $F_{\mu\nu}=F^{\alpha}{}_{\mu\alpha\nu}$, $F=F^\mu{}_\mu$,
$f_i$ ($i=1,2,\ldots,6$), $a_k$ ($k=1,2,3$) are indefinite
parameters, $f_0=(16\pi G)^{-1}$, $G$ is Newton's gravitational
constant (the light speed in the vacuum $c=1$). Gravitational
equations of PGTG are deduced from the action integral $ I = \int
{\left( {{\cal L}_g + {\cal L}_m } \right)\,}h d^4 x$, where
$h=\det{\left(h^i{}_\mu\right)}$ and ${\cal L}_m$ is the
Lagrangian of gravitating matter. Although the gravitational
Lagrangian (2) includes a number of indefinite parameters,
gravitational equations of PGTG for homogeneous isotropic models
(HIM) considered below depend weakly on the choice of quadratic
part of gravitational Lagrangian by virtue of their high spatial
symmetry. The investigation of HIM in the frame of PGTG leads to
some important physical consequences concerning the gravitational
interaction for usual gravitating matter.

\section{On gravitational interaction in PGTG}

The system of gravitational equations of PGTG corresponding to
gravitational Lagrangian (2) in general case is a complicated
system of differential equations. From mathematical point of view,
HIM are the most simple models, for which the system of
gravitational equations of PGTG has a sufficiently simple form.
From physical point of view, the study of HIM has important
cosmological applications. Isotropic cosmology including
inflationary cosmology in the frame of PGTG was built and
investigated in our works (see [21-26] and references herein). Now
we will discuss some physical consequences of general character
following from investigation of HIM. The dynamics of HIM in PGTG
is described in general case by three functions of time: the scale
factor of Robertson-Walker metrics $R(t)$ and two torsion
functions $S_{1}(t)$ and $S_{2}(t)$ determining the form of the
torsion tensor. The functions $S_{1}(t)$ and $S_{2}(t)$ have
different properties with respect to transformations of spacial
inversions: unlike $S_{1}(t)$ the function $S_{2}(t)$ has
pseudoscalar character. We will consider two types of HIM: HIM
with one torsion function $S_{1}$ ($S_{2}=0$) and HIM with two
torsion functions.

At first we will consider HIM with vanishing pseudoscalar torsion
function [21-23] filled by gravitating matter with energy density
$\rho$ and pressure $p$ (the average of spin distribution is equal
to zero). In this case gravitational equations of PGTG lead to the
following generalized cosmological Friedmann equations (GCFE):
\begin{equation}
\label{3}
\displaystyle{\frac{k}{R^2}+\left\{\frac{d}{dt}\ln\left[R\sqrt{\left|1+\alpha\left(\rho-
3p\right)\right|}\right]\right\}^2 }\displaystyle{ =\frac{8\pi
G}{3}\;\frac{\rho+
\frac{\alpha}{4}\left(\rho-3p\right)^2}{1+\alpha\left(\rho-3p\right)}
\, ,}
\end{equation}

\begin{equation}
\label{4}
\displaystyle{R^{-1}\,\frac{d}{dt}\left[\frac{dR}{dt}+R\frac{d}{dt}\left(\ln\sqrt{\left|1+\alpha\left(\rho
-
3p\right)\right|}\right)\right]} 
\displaystyle{=-\frac{4\pi G}{3}\;\frac{\rho+3p-
\frac{\alpha}{2}\left(\rho-3p\right)^2}{
1+\alpha\left(\rho-3p\right)}\, ,}
\end{equation}
where indefinite parameter $\displaystyle
\alpha=\frac{f}{3f_0\,^2}>0$ ($f = f_1  + \frac{{f_2 }} {2} + f_3
+ f_4  + f_5  + 3f_6$) has the inverse dimension of energy
density. The GCFE are obtained by using the expression for the
torsion function $S_{1}(t)$ following from gravitational equations
in the form:
\begin{equation}
\label{5} S_{1}=
-\frac{1}{4}\frac{d}{dt}\ln\left|1+\alpha(\rho-3p)\right|.
\end{equation}
Because the equations of motion for spinless matter in the frame
of PGTG (by minimal coupling with gravitation) have the same form
as in GR, the conservation law for gravitating matter has usual
form:
\begin{equation}\label{6}
    \dot \rho  + 3H\left( {\rho  + p} \right) = 0,
\end{equation}
where $H=\dot{R}/R $ is the Hubble parameter and a dot denotes the
differentiation with respect to time \footnote{It is easy to show
that the conservation law (6) follows directly from (3)-(4).}. The
difference of (3)--(4) from Friedmann cosmological equations of GR
is connected with terms containing the parameter $\alpha$. The
value of $\alpha^{-1}$ determines the scale of extremely high
energy densities. Solutions of GCFE coincide practically with
corresponding solutions of GR, if the energy density is small
$\left|\alpha(\rho-3p)\right|\ll 1$ ($p\neq\frac{1}{3}\rho$). The
difference between GR and PGTG can be significant at extremely
high energy densities $\left|\alpha(\rho- 3p)\right| \sim 1$,
where the dynamics of HIM depends essentially on space-time
torsion.

The structure of GCFE (3)-(4) ensures regular behavior of
cosmological solutions. It is because the gravitational
interaction at extreme conditions changes and has the repulsive
character [22]. Unlike the gravitational repulsion effect in the
frame of Einstein-Cartan theory, which appears for spinning matter
[27] and critically depends on the spin description [28], in the
case of discussed HIM the gravitational repulsion takes place for
usual spinless matter that leads to regularity of ordinary
cosmological HIM. In order to demonstrate this fact in the case of
inflationary cosmological models, we will consider HIM filled with
scalar field $\phi$ minimally coupled with gravitation and
gravitating matter with equation of state in the form
$p_m=p_m(\rho_m)$ (values of gravitating matter are denoted by
means of index "m"). Then the energy density $\rho$ and the
pressure $p$ take the form
\begin{equation}
\label{7} \rho=\frac{1}{2}\dot{\phi}^2+V+\rho_m \quad (\rho>0),
\quad p=\frac{1}{2}\dot{\phi}^2-V+p_m,
\end{equation}
where $V=V(\phi)$ is a scalar field potential. Because the energy
density $\rho$ is positive and $\alpha>0$, from equation (3) in
the case $k=+1$, $0$ follows the relation:
\begin{equation}
\label{8}
Z=1+\alpha\left(\rho-3p\right)=1+\alpha\left(4V-\dph^2+\rhm-3\prm\right)\ge
0.
\end{equation}
The relation (8) is valid also for cosmological solutions of open
type ($k=-1$)[21]. The domain of admissible values of scalar field
$\phi$, time derivative $\dot{\phi}$ and energy density $\rho_m$
determined by (8) is limited in space $P$ of these variables by
bound $L$ defined as
\begin{equation} \label{9}
Z=0\quad \mbox{or}\quad
\dot\phi=\pm\left(4V+\alpha^{-1}+\rhm-3\prm\right)^{\frac{1}{2}}.
\end{equation}
Unlike GR at compression stage the time derivative $\dot{\phi}$
does not diverge, and by reaching the bound $L$ the transition to
the second part of cosmological solution containing the expansion
stage takes place. From cosmological equation (3) by using the
conservation law (6) follows that in space $P$ there are extremum
surfaces, in points of which the Hubble parameter vanishes [21,
23]. Extremum surfaces play the role of "bounce surfaces", because
the time derivative of the Hubble parameter is positive on the
greatest part of these surfaces in the case of scalar field
potentials applying in chaotic inflation. All cosmological
solutions have bouncing character and are regular with respect to
metrics, Hubble parameter and its time derivative. If gravitating
matter satisfies standard conditions (energy density is positive,
energy dominance condition is valid), any cosmological solution is
not limited in the time, and before the expansion stage
cosmological solution contains the compression stage and regular
transition from compression to expansion. Note that the character
of gravitational interaction in the frame of PGTG depends
essentially on properties of gravitating matter and at first of
all on its equation of state. The effect of gravitational
repulsion at extreme conditions in considered HIM takes place by
virtue of the following condition for total energy density and
pressure: $p>\frac{1}{3}\rho$. In the case of inflationary HIM
including together with usual gravitating matter also scalar
fields this condition is realized at certain moment of compression
stage always independently on conditions for usual matter: $p_{m}
=\frac{1}{3}\rho_{m}$ or $p_{m}\ > \frac{1}{3}\rho_{m}$. For
gravitating matter the condition $p_{m}\ > \frac{1}{3}\rho_{m}$ is
valid also at sufficiently high energy densities [29]. The HIM
filled with a such matter without scalar fields have the limiting
energy density, which is reaching at a bounce and is determined
from the relation $Z=0$ [30]. During the expansion stage, when the
energy density becomes sufficiently small and the equation of
state changes ($p < \frac{1}{3}\rho$),  the GCFE lead to
additional gravitational attraction in comparison with GR and
Newton's theory of gravity. In particular, at matter dominating
stage with equation of state for dust ($p=0$), by taking into
account the relation $\left(\alpha\rho\right) \ll 1$, it is easy
to obtain from (3)--(4) in the case $k=0$ that
\begin{equation}
\label{10} \ddot {R}/R = -\frac{4\pi G}{3} \rho (1 + 9 \alpha
\rho).
\end{equation}
According to (10) the force of gravitational attraction is
$(1+9\alpha\rho)$ times bigger than in Newton's theory of gravity.

The space-time torsion in PGTG can lead to gravitational repulsion
effect not only at extreme conditions, but also at very small
energy densities. Such situation takes place in the case of HIM
with two torsion functions [24, 26]. Cosmological equations for
such HIM include also the pseudoscalar torsion function $S_{2}$
with its first time derivative and contain besides $\alpha$ two
others indefinite parameters: $b = a_2 - a_1$ with dimension of
parameter $f_0$ and dimensionless parameter $\veps$, which is the
function of coefficients $f_i$. If $|\veps|\ll 1$, the
pseudoscalar torsion function contains at asymptotics, where
physical parameters of cosmological model are sufficiently small,
some not vanishing value and is equal to:
\begin{equation}\label{11}
S_2^2  = \frac{1}{12b} \left[ \rho - 3p + \alpha^{-1}(1 - b/f_0)
\right].
\end{equation}
As a result cosmological equations at asymptotics take the form of
cosmological Friedmann equations with effective cosmological
constant induced by pseudoscalar torsion function:
\begin{equation}\label{12}
    \frac{k} {{R^2 }} + H^2  = \frac{1} {{6b }}\left[ \rho  + \frac{1}{4} \alpha^{-1}(1 - b/f_0)^2 \right],
\end{equation}
\begin{equation}\label{13}
    \dot H + H^2  =  - \frac{1} {{12b }}\left[ \rho  + 3p - \frac{1}{2} \alpha^{-1}(1 - b/f_0)^2 \right].
\end{equation}
By using at asymptotics the equation of state for dust ($p=0$), we
see that cosmological equations (12)-(13) lead to observable
accelerating cosmological expansion by certain relation between
indefinite parameters $b$ and $\alpha$. If we suppose that the
scale of extremely high energy densities defined by $\alpha^{-1}$
is larger than the energy density for quark-gluon matter, but less
than the Planckian one, then we can obtain the corresponding
estimation for $b$, which is very close to $f_0$. Hence, the
acceleration of cosmological expansion in PGTG has geometrical
nature and is connected with the change of gravitational
interaction induced by space-time torsion. The investigation of
inflationary HIM with two torsion functions at extreme conditions
at the beginning of cosmological expansion shows that the PGTG
allows to build totally regular inflationary Big Bang scenario by
classical description of gravitational field. If the energy
density and values of torsion functions at transition stage from
compression to expansion are less than the Planckian ones, quantum
gravitational era was absent by evolution of the Universe. If the
Planckian conditions were realized at the beginning of
cosmological expansion, quantum gravitational corrections have to
be taken into account; however, classical cosmological equations
of PGTG ensure the regular character of the Universe evolution.

\section{Conclusion}

From our consideration given above follows that the PGTG can have
the important meaning for theory of gravitational interaction. The
PGTG leads to certain principal differences in  comparison with GR
concerning the character of gravitational interaction for usual
spinless gravitating matter. According to PGTG, the domain of
applicability of GR is limited, namely in the case of cosmological
HIM the domain of admissible energy densities has upper limit
determined by $\alpha^{-1}$ and lower limit equal to $\frac{1}{4}
\alpha^{-1}(1 - b/f_0)^2$. The investigation of HIM with two
torsion functions shows that the torsion can be important in
Newtonian approximation (see (11)) and the Newton's law of
gravitational attraction has limits of its applicability in the
case of usual gravitating systems with sufficiently small energy
densities. As it was noted above, the law of gravitational
interaction for such gravitating systems can include corrections
corresponding to additional attraction. This means that the
investigation of not homogeneous gravitating systems at galactic
scales, in particular, of spherically-symmetric systems in the
frame of PGTG is of direct physical interest in connection with
the problem of dark matter of GR with the purpose of its solution.

\end{document}